\documentclass[aps,
		prd,
		reprint,
		twocolumn,
		superscriptaddress,
		shortbibliography,
		nofootinbib,
		floatfix,
		notitlepage
		]{revtex4}
\pdfoutput=1
     
\usepackage{graphicx}

\usepackage{amsmath}
\usepackage{amssymb}
\usepackage{mathrsfs}
\usepackage{bbm}
\usepackage{bbold}
\usepackage{amsfonts}

\usepackage[usenames,dvipsnames]{xcolor}
\usepackage{slashed}

\newcommand{\vphi}{\varphi}
\newcommand{\eps}{\varepsilon}

\newcommand{\trm}[1]{\textrm{#1}}
\newcommand{\tsf}[1]{\textsf{#1}}
\newcommand{\tr}{\textrm{tr}\,}

\newcommand{\be}{\begin{equation}}
\newcommand{\ee}{\end{equation}}
\newcommand{\bea}{\begin{eqnarray}}
\newcommand{\eea}{\end{eqnarray}}
\newcommand{\bi}{\begin{itemize}}
\newcommand{\ei}{\end{itemize}}

\newcommand{\Ai}{\trm{Ai}}
\newcommand{\Gi}{\trm{Gi}}

\newcommand{\nn}{\nonumber}

\newcommand{\bra}[1]{\langle #1 |}
\newcommand{\ket}[1]{|#1\rangle}
\newcommand{\braket}[2]{\langle #1|#2\rangle}

\newcommand{\av}[1]{\langle#1\rangle}
\newcommand{\drm}{\mathrm{d}}

\newcommand{\figref}[1]{Fig. \ref{#1}}

\newcommand{\eqnref}[1]{Eq. (\ref{#1})}

\newcommand{\tabref}[1]{Tab.$~$\ref{#1}}
\newcommand{\vkap}{\varkappa}

\begin{document}

\title{Probing the vacuum as a chiral medium}

\author{T.~Heinzl}
\email{t.heinzl@plymouth.ac.uk}
\affiliation{Centre for Mathematical Sciences, Plymouth University, Plymouth, PL4 8AA, United 
Kingdom}
\author{B.~King}
\email{b.king@plymouth.ac.uk}
\affiliation{Centre for Mathematical Sciences, Plymouth University, Plymouth, PL4 8AA, United 
Kingdom}
\author{A.~Mercuri-Baron}
\email{anthony.mercuri-baron@plymouth.ac.uk}
\affiliation{Centre for Mathematical Sciences, Plymouth University, Plymouth, PL4 8AA, United 
Kingdom}

\date{\today}

\begin{abstract}
We study the \emph{circular birefringence} experienced by linearly polarised photons colliding with a circularly polarised background creating a vacuum of definite chirality (handedness). For this scenario the standard Heisenberg-Euler approach fails and must be supplemented by derivative corrections which we match to known Hilbert series. Choosing a plane wave background, we find equivalence between three approaches: (i) adding derivative corrections to the Heisenberg-Euler Lagrangian; (ii) improving the locally constant field approximation to the one-loop polarisation tensor; (iii) performing a low-energy expansion of the direct $2\to 2$ QED photon-photon scattering amplitude. Going beyond plane-wave backgrounds, we analyse an example of a circularly polarised standing wave sensitive to derivative corrections. We find a parameter regime where these corrections could be probed in experiments.
\end{abstract}

\maketitle

\section{Introduction}
Vacuum birefringence is a long-standing prediction of quantum electrodynamics (QED), which still remains to be tested in intense fields. Its history and current status has recently been reviewed in \cite{Ahmadiniaz:2024xob}. We thus only re-emphasise that despite tantalising hints from astrophysics \cite{Mignani:2016fwz}, photon-scattering experiments at colliders \cite{ATLAS:2017fur,CMS:2018erd,ATLAS:2019azn,Brandenburg:2022tna} and long-running dedicated cavity experiments \cite{Agil:2021fiq,Ejlli:2020yhk}, vacuum birefringence has so far remained elusive. However, an experimental confirmation is actively pursued using the high intensity of powerful optical laser pulses \cite{schlenvoigt15,Ahmadiniaz:2024xob,Rinderknecht:2025zgu} to coherently enhance the very small cross-section \cite{Heinzl:2024cia} for vacuum birefringence. (For reviews on photon-photon scattering and strong-field QED see, e.g.~\cite{Marklund.RevModPhys.78.591.2006,DiPiazza.RevModPhys.84.1177.2012,King:2015tba,Fedotov:2022ely,Sarri:2025qng}.)

The leading contribution to vacuum birefringence is from a single electron-positron loop, dressed by the electromagnetic background. For photon-laser and laser-laser collisions, the centre-of-mass energy is typically much less than the electron rest mass. In this case, an effective approach can be used to calculate experimental observables, based on an effective Lagrangian which schematically can be written as
\be \label{L.EFF0}
  \mathcal{L}_{\tsf{eff}} = \sum_{n,i} \mathcal{L}_{n,i} \; , \quad 
  \mathcal{L}_{n,i} \sim  c_{n,i}\partial^{(i)} \mathscr{O}^{(n)} \; .
\ee
Here $c_{n,i}$ are low-energy constants with the integer $n$ denoting the number of electromagnetic field tensors, $F$, appearing in the operator $\mathscr{O}^{(n)}$ and the index $i$ counting the number of derivatives\footnote{For a given number of derivatives and field strengths, there are typically several inequivalent terms, $\mathcal{L}_{n,i,j}$, each with its own coupling, $c_{n,i,j}$, $j = 1, 2 , \ldots, k$. To avoid index inflation, we suppress the subscript $j$.}acting on $\mathscr{O}^{(n)}$. The many scenarios considered in the literature until now (as reviewed in \cite{Ahmadiniaz:2024xob}), study vacuum \emph{linear} birefringence, for which the leading order (LO) contribution stems from $\mathcal{L}_{4,0}$, i.e.\ from the terms without extra derivatives. However, for vacuum \emph{circular} birefringence \cite{Affleck:1987gf,King:2023eeo}, which can arise in a circularly-polarised (CP) laser, we will show the leading contribution is from the next-to-leading order (NLO) (in derivatives), i.e. from $\mathcal{L}_{4,2}$. Circular birefringence therefore provides an experimental test of higher-dimensional operators in the effective field theory.

We begin in Sec.~II by reviewing derivative corrections of the form $\mathcal{L}_{4,n}$ and show how they are related to helicity amplitudes expressed through Mandelstam variables. In Sec.~III, we introduce our set-up and an intuitive formalism based on the language of two-level systems (Jones vectors). In Section~IV, we focus on the explicit example of a plane wave background, and demonstrate the equivalence of three approaches to calculate the helicity flip probabilities: (i) adding derivative corrections to the locally constant field approximation (LCFA), (ii) adding derivative corrections to the HE Lagrangian and (iii) evaluating the low-energy limit of QED 4-photon amplitudes. In Sec.~V we use the effective approach to calculate circular birefringence in a scenario replacing the highly symmetric plane wave by a standing circular wave. Sec.~VI presents our conclusions.

\section{Derivative Corrections to low intensity Heisenberg-Euler Lagrangian}

At energies below the pair threshold, QED (in $d=4$, which we consider throughout) is well described by augmenting the Maxwell Lagrangian with the LO Heisenberg-Euler (HE) correction \cite{Heisenberg:1936nmg},
\bea \label{HE.LO}
   \mathcal{L}_{4,0} &\equiv& \frac{\alpha^2}{360 m^4} 
   \left[ 4 (\tr F^2)^2 + 7 (\tr F \tilde{F})^2 \right] \nonumber \\
   &\equiv& 
   \frac{\alpha^2}{180 m^4} \left[ -5 (\tr F^2)^2 
   + 14  \, \tr F^4 \right]  \; .
\eea
Note that there are \emph{two} independent terms at this order.
In accordance with (\ref{L.EFF0}), the subscripts $4,0$ signal that each term in the LO HE Lagrangian contains $n=4$ factors of the Maxwell field strength, $F$, and no derivatives. For the second equality, we have used the identity \cite{Davila:2013wba},
\be 
  \big( \tr F \tilde{F} \big)^2 = 4 \, \tr F^4 - 
  2 \, \big( \tr F^2 \big)^2 \; ,
\ee
which allows one to write $\mathcal{L}_{4,0}$ entirely in terms of the field strength, $F$, without employing its dual, $\tilde{F}$, with components $\tilde{F}^{\mu\nu} = (1/2) \epsilon^{\mu\nu\rho\sigma} F_{\rho\sigma}$. The Lagrangian (\ref{HE.LO}), first considered by Heisenberg's students Euler and Kockel \cite{Euler:1935zz}, is the leading term in a weak-field expansion that is known to all orders in $F$ and $\tilde{F}$ \cite{Heisenberg:1936nmg}, hence in resummed form. Due to Lorentz and gauge invariance, the complete HE Lagrangian only depends on the scalar and pseudo-scalar  invariants 
\bea
  \mathcal{S} &=& - \frac{1}{4} F^{\mu\nu} F_{\mu\nu} 
  \equiv \frac{1}{4} \, \tr F^2 \; , \label{S} \\
  \mathcal{P} &=& - \frac{1}{4} F^{\mu\nu} \tilde{F}_{\mu\nu} \equiv
  \frac{1}{4} \, \tr  F\tilde{F} \; . \label{P}
\eea
(In what follows, we will often omit the trace symbol to save space.)
As QED conserves parity, only even powers of $\mathcal{P}$ (or $\tilde{F}$) are present. Allowing for parity violation would induce odd powers of $\mathcal{P}$ and hence enlarge the operator basis. At LO, for instance, a third term, proportional to $\mathcal{SP}$, would appear.

Physically, the LO Lagrangian (\ref{HE.LO}) describes the self-interactions of photons, hence photon-photon scattering, induced by the QED one-loop box diagram. At low centre-of-mass energy squared, $s \ll m^2$, the fermion loop is replaced by an effective vertex of order $\alpha^2/m^4$. At NLO in field strength corresponding to $\mathcal{L}_{6,0} $, the effective vertex is order $\alpha^3/m^8$, and so on.

From a modern perspective, the HE Lagrangian is a paradigmatic example of a low-energy effective field theory: the effective Lagrangian is an expansion in terms of operators of increasing mass dimension, each term weighted by a dimensionless coupling constant divided by the appropriate energy or cut-off scale (here the electron mass, $m$). In condensed notation, this expansion (in $d=4$) may be written as in (\ref{L.EFF0})
\be \label{L.EFF}
  \mathcal{L}_\mathrm{eff} =\sum_{n,i} \mathcal{L}_{n,i} \equiv 
  \sum_{n,i} c_{n,i} \frac{\alpha^{n/2}}{m^{2n + i - 4}} 
  \partial^{(i)} F^{n} \; ,
\ee
guaranteeing mass dimension 4 for the Lagrangian with dimensionless low energy constants $c_{n,i}$. Furry's theorem demands that the number $n$ of field operators $F$ (or its dual) is even. Their total mass dimension is $2n$ to which one has to add an even number $i$ counting the derivatives as well as their mass dimension. The original all-orders HE Lagrangian is the sum of all terms with $n \ge 4$ and no derivatives ($i=0$), which hence represents the QED effective action in a constant background. The low-energy couplings, $c_{n,i}$, are determined by the QED Lagrangian, the ultra-violet completion of the effective Lagrangian (\ref{L.EFF}). The underlying procedure is referred to as matching: one demands that the low-energy limit of the QED photon scattering amplitude agrees with its pendant calculated in the effective theory given by $\mathcal{L}_\mathrm{eff}$.  

In this paper, we are interested in the derivative corrections with $i=2$, the minimum value required by Lorentz invariance. By construction, these corrections extend the validity of the effective Lagrangian to higher energies. At LO in field strength, $n=2$, there is only one term, the LO vacuum polarisation correction, which has been known since the original calculations of Dirac \cite{Dirac:1934} and Heisenberg \cite{Heisenberg:1934pza} at the dawn of QED. In our notation, one has
\be
  \mathcal{L}_{2,2} = \frac{1}{60\pi} \frac{\alpha}{m^2} F_{\mu\nu} \Box 
  F^{\mu\nu} 
  \; .
\ee
This is often referred to as the (low-energy limit of) the Uehling potential \cite{Uehling:1935uj} induced by vacuum polarisation. As shown in \cite{Kong:1998ic}, this term is a redundant operator in the effective field theory as it can be removed by a field redefinition. Adding just $\mathcal{L}_{2,2}$ to the Maxwell Lagrangian (with $c_{2,2}$ as a free parameter), yields what is called Bopp-Podolsky electrodynamics \cite{Bopp:1940,Podolsky:1942zz}. It was originally introduced to modify the UV behaviour of the theory and to deal with the problem of radiation reaction. This has recently been taken up again in \cite{Zayats:2013ioa,Gratus:2015bea}.

The next correction with two derivatives involves four field strengths ($n=4$). To the best of our knowledge, it has first been determined by Dicus et al. \cite{Dicus:1997ax} who find \emph{three} terms at this order,
\begin{align} \label{L.4.2}
  \mathcal{L}_{4,2} = \frac{1}{945}\frac{\alpha^2}{m^6} &
  \bigg[ (\partial_\alpha \partial_\beta F_{\mu\nu}) F^{\mu\nu}  
  F^\alpha_{\;\;\;\rho} F^{\rho \beta} \nonumber \\
  &+  3 (\partial^\alpha F_{\mu\nu} ) (\partial_\alpha F^{\mu\nu} ) 
  F_{\rho\sigma} F^{\rho\sigma}  \nonumber \\
  &+ 11 (\partial^\alpha F_{\mu\nu}) F^{\nu\rho} 
  (\partial_\alpha F_{\rho\sigma}) F^{\sigma\mu} 
  \bigg] \; .
\end{align}
This correction has been recalculated at least twice, first by Gusynin and Shovkovy, who obtain seven terms altogether \cite{Gusynin:1998bt}. However, they state that on-shell (when $k^2 = 0$, equivalent to using the free equation of motion), their answer reduces to the three terms in (\ref{L.4.2}). More recently, a result with four terms has been reported by Karbstein \cite{Karbstein:2021obd}. It can be straightforwardly shown however, that two of Karbstein's terms can be combined after an integration by parts. One can check the three couplings in \eqnref{L.4.2} by leaving them undetermined and matching with the QED amplitudes for photon-photon scattering in the low energy limit \cite{DeTollis:1964una}. We performed this check and confirm the couplings in  \eqnref{L.4.2} as given by \cite{Dicus:1997ax}. For the sake of completeness, we note that Dicus et al.\ have also worked out the next order ($n = i = 4$), which again yields three terms,
\begin{align} \label{L.4.4}
  \mathcal{L}_{4,4} = \frac{1}{9450}\frac{\alpha^2}{m^8} &
  \bigg[ -33 (\partial_\alpha F_{\mu\nu}) (\partial^\alpha F^{\mu\nu}) 
  (\partial_\beta F_{\rho\sigma})  (\partial^\beta F^{\rho\sigma})  
  \nonumber \\
  & + 106 (\partial_\alpha F_{\mu\nu}) (\partial^\beta F^{\mu\nu}) 
  (\partial^\alpha F_{\rho\sigma})  (\partial_\beta F^{\rho\sigma}) 
  \nonumber \\
  & - 262 (\partial^\alpha \partial_\beta F_{\mu\rho}) F^{\rho\nu} 
  (\partial^\mu \partial_\nu F_{\alpha\sigma}) F^{\sigma\beta} \bigg] \; .
\end{align} 
The results above imply that at $n=4$ the number of terms with $i = 0,2,4$ derivatives in the effective Lagrangian equals $k=2,3,3$, respectively. We have seen, however, that even for $i=2$ there are three different answers in the literature, although these could be boiled down to a single Lagrangian with $k = 3$ terms. It would be useful to have a general counting scheme for these operators, in particular when going to higher orders. Fortunately, such a counting scheme exists and goes by the name of \emph{Hilbert series}. Put in simplest terms, a Hilbert series is just the generating function for the number $k$ of operators of a given mass dimension in an effective field theory. As usual, cf.\ (\ref{L.EFF0}) and (\ref{L.EFF}), these operators are ordered by increasing mass dimension. For any given mass dimension, the choice of operators is not unique as there are a number of redundancies such as integration-by-parts (IBP) and equation-of-motion (EOM) ambiguities which allow for field redefinitions. This freedom of choice is accompanied by restrictions due to symmetry. For instance, if one imposes parity symmetry, the number of operators gets reduced as briefly discussed above. Analogous statements hold for other symmetries, both local and global. In any case, the possible equivalent choices of operators yield the same physics (or S-matrix). The existence of a Hilbert series then implies that the \emph{number} of operators (hence the number of couplings) for a given mass dimension is unique. The literature on Hilbert series by now is too extensive for providing a complete list of references. Brief reviews with an emphasis on effective field theory may be found e.g.\ in \cite{Lehman:2015via,Lehman:2015coa,Henning:2015alf,Henning:2015daa,Henning:2017fpj} which also list some relevant earlier references.

For the case at hand, i.e., the fourth order field strength operators contributing to $\mathcal{L}_{4,i}$, the Hilbert series may be found in Table~2 of \cite{Henning:2017fpj} (the case including parity symmetry), and explicitly reads
\bea \label{HILBERT.P}
  H_{F^4}^P (t) &=& \frac{2 + 3t^2  + t^4 - t^6}{(1-t^4)(1 - t^6)} 
  \nonumber \\
   &=&  2 + 3t^2 + 3t^4 + 4t^6 + 6t^8 + 5t^{10} 
   + \ldots \; . 
\eea 
Looking at the expansion coefficients in the second line we conclude that there are 2 terms without derivatives (LO HE), cf.\ (\ref{HE.LO}), 3 terms with 2 derivatives, cf.\ (\ref{L.4.2}), 3 terms with 4 derivatives, cf.\ (\ref{L.4.4}), 4 terms with 6 derivatives, and so on.

For later use we also quote the result without parity symmetry from \cite{Henning:2017fpj},
\bea \label{HILBERT.NP}
  H_{F^4}^{\slashed{P}} (t) &=& 
  \frac{3 + 5t^2 + t^4 - 2t^6}{(1-t^4)(1 - t^6)} 
  \nonumber \\
   &=&  3 + 5t^2  + 4t^4  + 6t^6  + 9t^8  + 7t^{10}  
   + \ldots \; ,
\eea
which unsurprisingly shows an increase in the number of operators. As expected, at lowest order (no derivatives, $i=0$), there is one additional term, which we identified earlier as $\mathcal{S}\mathcal{P}$, a pseudo-scalar that indeed violates parity. QED of course predicts that its coupling is zero, but an axion field, for instance, would couple to this operator. At NLO (two derivatives, $i=2$) the number of operators increases from 3 to 5. This has recently been confirmed in \cite{Ruhdorfer:2019qmk}.

There is a rather useful way of addressing the issue of parity by introducing the `Silberstein variables' \cite{Silberstein:1907}, i.e., the linear combinations,
\be \label{F.PM}
  F_\pm := F \pm i \tilde{F} \; .
\ee
These are (anti)-self-dual and transform under parity according to $F_\pm \to F_\mp$. For this reason, they are also referred to as chiral fields---the subscripts denoting their chirality ($\pm = $ left/right). The tensors $F_\pm$ have a number of useful properties. For instance, they allow for an elegant classification of electro-magnetic field strength tensors according to their Lorentz properties \cite{Stephani:2004ud}. For our purposes they provide an alternative choice of operator basis which has been employed in a similar context before in \cite{Colwell:2015wna,Ruhdorfer:2019qmk}.

To actually perform the change of basis we trade $F$ and $\tilde{F}$ for $F_\pm$ in the invariants (\ref{S}) and (\ref{P}), 
\bea
  \mathcal{S} &=& \frac{1}{16} (\tr F_+^2 + \tr F_-^2) \; , \label{S.PM}\\
  \mathcal{P} &=& \frac{1}{16i} (\tr F_+^2 - \tr F_-^2) \label{P.PM} \; .
\eea
Under parity, the labels $+$ and $-$ get swapped such that $\mathcal{P}$ changes sign as befits a pseudo-scalar. Plugging (\ref{S.PM}) and (\ref{P.PM}) into (\ref{HE.LO}) and omitting the trace symbols, the LO HE Lagrangian becomes \cite{Colwell:2015wna}
\be \label{L.40.PM}
  \mathcal{L}_{4,0} = \frac{\alpha^2}{5760 \, m^4} \left\{ -3 \left[ 
  (F_+^2)^2 +   (F_-^2)^2 \right] + 22 \;  F_+^2 \,  F_-^2
  \right\} \; .
\ee
We note that parity conservation requires that the two terms $ (F_\pm^2)^2$ come with the same pre-factor, so that the overall number of independent terms remains two. If one allows for parity violation, this is no longer true, and one has three independent terms as stated above, cf.\ \cite{Ruhdorfer:2019qmk}.

Recasting $\mathcal{L}_{4,2}$ from (\ref{L.4.2}) into chiral form is a more onerous task. From (\ref{HILBERT.NP}) we know that there are altogether 5 operators accompanied by two derivatives. These operators are $F_\pm^4$, $F_+^2 F_-^2$ which are already present in $\mathcal{L}_{4,0}$, see (\ref{L.40.PM}), plus two additional ones, namely $F_\pm^3 F_\mp$ \cite{Ruhdorfer:2019qmk}. In the parity symmetric case (QED), quartic and cubic terms appear in the symmetric combinations $F_+^4 + F_-^4$ and  $F_+^3 F_- + F_-^3 F_+$ which reduces the number of independent operators to 3 as required by the associated Hilbert series (\ref{HILBERT.P}). A somewhat tedious calculation utilising IBP and EOM redundancies yields the explicit chiral representation,
\be
  \mathcal{L}_{4,2} = \frac{\alpha^2}{15120 m^6}
  \left\{ \mathcal{O}^{++++} + O^{++--} + 
  O^{+++-} + ({\scriptstyle + \leftrightarrow -})  \right\},
\ee
with the chiral operators given by
\begin{align}
  O^{\pm\pm\pm\pm} &\sim -13 (F_\pm \partial F_\pm)^2 
  - 4 F_\pm (F_\pm^2 \partial^2) F_\pm \, , \\
  O^{\pm\pm\mp\mp} &\sim 9 (F_\pm \partial F_\pm) 
  (F_\mp \partial F_\mp) 
  + 4  F_\mp (F_\pm^2 \partial^2) F_\mp , \\
  O^{\pm\pm\pm\mp} &\sim 4 F_\pm (F_\pm F_\mp + F_\mp F_\pm)
  \partial^2 F_\pm \; ,
\end{align}
suppressing the coupling constants throughout. In the above, we have used the abbreviation (omitting $\pm$ subscripts on $F$),
\be 
  F^2 \partial^2 \equiv F^\alpha_{\;\;\;\lambda} 
  F^{\lambda\beta} \partial_\alpha \partial_\beta \; .
\ee
Once the effective Lagrangian has been written in terms of the chiral fields, one can employ the one-to-one correspondence between the basic operators and the helicity amplitudes for photon-photon scattering. Informally, this can be written as
\be \label{OP.AMP}
  F_{\lambda_1} F_{\lambda_2} F_{\lambda_3} F_{\lambda_4} \sim
  M^{\lambda_1 \lambda_2 \lambda_3 \lambda_4} \; ,
\ee
where each $\lambda_i = \pm$ represents a helicity label. The correspondence has some interesting consequences. First note that the LO Lagrangian (\ref{L.40.PM}) does not contain terms with an odd number of $F_\pm$s. Hence, the associated helicity amplitudes vanish \cite{Colwell:2015wna},
\be \label{M.ODD}
  M^{\pm\mp\mp\mp} = M^{\mp\pm\mp\mp} = M^{\mp\mp\pm\mp} 
  = M^{\mp\mp\mp\pm} = 0 \; .
\ee
At the root of this property is the fact that the invariants $\mathcal{S}$ and $\mathcal{P}$ contain only even powers of $F_\pm$, cf.\ (\ref{S.PM}) and (\ref{P.PM}). As this is true to all orders in these invariants, all amplitudes with an odd number of either $+$ or $-$ helicities vanish for the complete HE Lagrangian, $\mathcal{L}_\mathrm{HE} \equiv \sum_n \mathcal{L}_{n,0}$. In other words, at LO in the derivative expansion ($i=0$) resummed over all powers $n$ of field strength, $F$,  (\ref{M.ODD}) remains valid \cite{Martin:2003gb}. For $n=4$, $i=0$, the non-vanishing LO helicity amplitudes are \cite{DeTollis:1965vna,Colwell:2015wna,AH:2023ewe}
\bea \label{M.LO}
  M_\mathrm{LO}^{\pm\pm\mp\mp} &=& 
  \frac{11 \alpha^2}{45 m^4} s^2 \; , 
  \nonumber \\
  M_\mathrm{LO}^{\pm\mp\pm\mp} &=& 
  \frac{11 \alpha^2}{45 m^4} t^2 \; , 
  \nonumber \\
  M_\mathrm{LO}^{\pm\mp\mp\pm} &=& 
  \frac{11 \alpha^2}{45 m^4} u^2 , 
  \nonumber \\
  M_\mathrm{LO}^{\pm\pm\pm\pm} &=& 
  - \frac{\alpha^2}{15 m^4} 
  (s^2 + t^2 + u^2) 
  \; ,
\eea
where all photons are assumed incoming and $s,t,u$ are the usual Mandelstam invariants. The first three lines represent the operator $F_+^2 F_-^2$, the last line the operator $F_+^4 + F_-^4$. We note in passing that these amplitudes have recently been calculated to two-loop order in QED and QCD \cite{AH:2023ewe,AH:2023kor} and thus are known to order $\alpha^3$ and $\alpha^2 \alpha_s$ for arbitrary centre-of-mass energy ($\alpha_s$ denoting the strong coupling constant). For our purposes, the first few orders in the low-energy/derivative expansion are sufficient.

At NLO in derivatives ($n=4$, $i=2$) one first encounters the new operator $F_+^3 F_- + F_-^3 F_+$. In consequence, the amplitudes (\ref{M.ODD}) no longer vanish \cite{DeTollis:1965vna,Colwell:2015wna},
\be 
   M_\mathrm{NLO}^{\pm\mp\mp\mp} 
   = M_\mathrm{NLO}^{\mp\pm\mp\mp} 
   = M_\mathrm{NLO}^{\mp\mp\pm\mp} 
   = M_\mathrm{NLO}^{\mp\mp\mp\pm} 
   = \frac{-\alpha^2}{315 m^6} stu. 
\ee
These are \emph{cubic} in the Mandelstam variables as $stu = -(s^3 + t^3 + u^3)/3$, a consequence of $s + t + u = 0$. The NLO corrections to the amplitudes (\ref{M.LO}) are found to be \cite{DeTollis:1965vna,Colwell:2015wna}
\bea \label{M.NLO}
  M_\mathrm{NLO}^{\pm\pm\mp\mp} &=& 
  -\frac{4 \alpha^2}{315 m^6} s^3 \; , 
  \nonumber \\
  M_\mathrm{NLO}^{\pm\mp\pm\mp} &=& 
  -\frac{4 \alpha^2}{315 m^6} t^3 \; , 
  \nonumber \\
  M_\mathrm{NLO}^{\pm\mp\mp\pm} &=& 
  -\frac{4 \alpha^2}{315 m^6} u^3 \; , 
  \nonumber \\
  M_\mathrm{NLO}^{\pm\pm\pm\pm} &=& \frac{2\alpha^2}{63 m^6} 
  stu \; .
\eea
These NLO contributions are normally sub-leading since the  Mandelstam variables are small compared to the electron rest mass (squared) in scenarios considered suitable for laser experiments. However, in the case of \emph{circular birefringence}, the LO terms are suppressed and the NLO terms are dominant as will be shown below. Before addressing this issue, we introduce our conventions and some intuitive formalism essentially borrowed from optics.

\section{Set-up and formalism}

\begin{figure}[h!!]
\centering
\includegraphics[width=8cm]{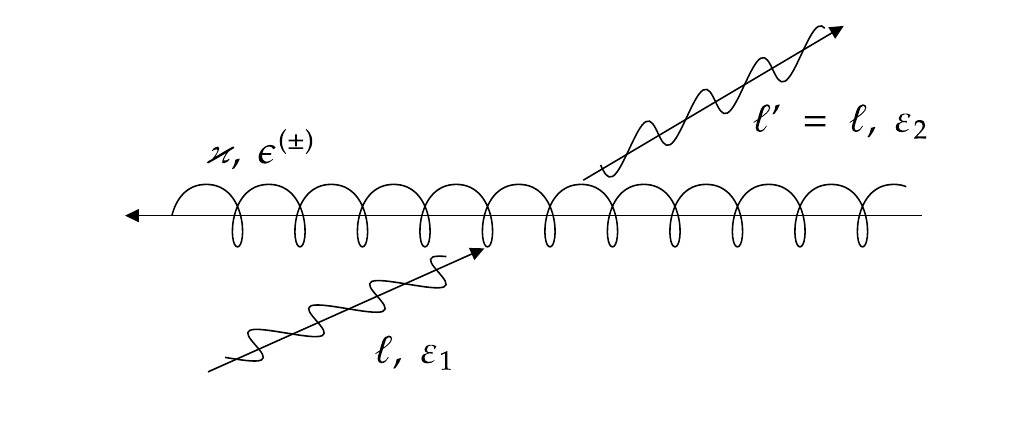}
\caption{Microscopic view of the circular birefringence scenario: There is a non-vanishing probability for the \emph{linear} polarisation of probe photons to flip upon scattering off the CP plane-wave background (momentum and polarisation labels shown).} \label{fig:sketch1}
\end{figure}

Let us consider the scattering of a linearly polarised (LP) photon in a CP  plane wave electromagnetic background, characterised by fixed light-like 4-momentum, $\vkap$, $\vkap^2 = 0$. The situation is sketched in \figref{fig:sketch1}. Modelling the intense laser background as a plane wave is a reasonable first approximation and allows for a straightforward demonstration of the main phenomenology. 

Mathematically, the background is described by a vector potential $a=eA$, rescaled with elementary charge $e>0$ and decomposed into positive and negative frequency components, $a = a^{(+)} + a^{(-)}$, according to
\bea
  a^{(\pm)} = \frac{m\xi}{\sqrt{2}} \epsilon^{(\mp)} \psi(\vphi) 
  \mbox{e}^{\pm i\vphi};\quad \epsilon^{(\mp)} = \frac{1}{\sqrt{2}}
  \left[\epsilon_{1} \mp i \epsilon_{2}\right] \; . \label{eqn:bg1}
\eea
The background phase is $\vphi=\vkap \cdot x$, while $\epsilon_{1,2}$ are orthonormal polarisation vectors normalised to $(\epsilon_{1,2})^{2}=-1$. It follows that $\epsilon^{(\mp)}\cdot \epsilon^{(\mp)\,\ast} = -1$ and $a\cdot a = -m^{2}\xi^{2}\psi^{2}(\vphi)$ which identifies $\xi$ as the dimensionless laser amplitude and $\psi$ as the phase dependent profile function. We choose the linear polarisation vectors of the probe photon to be
\bea 
  \eps_{j} = \epsilon_{j} - \vkap \,\frac{\ell \cdot \epsilon_{j}}{\vkap 
  \cdot \ell}, \label{eqn:photonPol1}
\eea
where $j \in \{1,2\}$ and $\ell \cdot \epsilon_j = 0$. 

A plane wave background is special because, due to light-front momentum conservation and the on-shell condition, the scattered photon momentum $\ell'$ is the same as the initial photon momentum $\ell$ (forward scattering). As there is no momentum transfer, the only quantum number that may change upon scattering is helicity or polarisation. We hence focus on the probability for polarisation flip of an incoming LP photon which we write as
\be \label{P.FLIP}
  \tsf{P}_{21} = |\mathfrak{M}^{(21)}|^{2}; \qquad 
  \mathfrak{M}^{(21)} =  \bra{2} \tsf{S} \ket{1} \equiv \tsf{S}_{21}, 
\ee
where $\ket{1}$ and $\ket{2}$ represent orthonormal photon linear polarisation states corresponding to \eqref{eqn:photonPol1}, $\mathfrak{M}$ is the amplitude and $\tsf{S}$ is the scattering matrix. In essence, for a photon colliding with a plane wave background, the polarisation state is described by a two-level system.

We hence employ the simplest possible Hilbert space, $\mathbb{C}^2$, with (polarisation) basis vectors $\ket{1}$ and $\ket{2}$ obeying the completeness relation $\mathbb{I} = \ket{1} \bra{1} + \ket{2}\bra{2}$. For illustration purposes, we consider a vacuum with homogeneous refractive index. The associated formalism is the starting point of, and nicely explained in, Baym's text on quantum mechanics \cite{Baym:1990} where further details (such as the mapping between polarisation states and the electromagnetic field) can be found. 

The S-matrix is just a unitary 2-by-2 matrix, its eigenvalues given by phases, $\exp{\left(i \delta_j\right)	}$, $j \in \{1,2\}$. Let us first assume an LP background which represents the standard scenario for vacuum birefringence (as recently reviewed in \cite{Ahmadiniaz:2024xob}). In this case, the S-matrix is indeed diagonal, 
\be\label{S.DIAG.1}
  \tsf{S}_{\mathrm{LP}} \equiv \sum_{i = 1,2} \exp(i \delta_i) 
  \ket{i}\bra{i} =
  \begin{pmatrix}
  \exp{i \delta_1} & 0 \\
  0 & \exp{i \delta_2}  
  \end{pmatrix} 
  \equiv \exp i \tsf{D}
  \; .
\ee
The matrix $\tsf{D} \equiv \mbox{diag}(\delta_1, \delta_2)$ defines the exponential representation of the S-matrix, $\tsf{S}_{\mathrm{LP}}$, a generalisation of which has recently been employed in the context of gravitational scattering amplitudes and their classical limit \cite{Damgaard:2021ipf,Damgaard:2023ttc}. It is useful to introduce half the phase sum and difference,
\be 
  \Sigma \equiv \frac{\delta_1 + \delta_2}{2} \; , \quad 
  \Delta \equiv \frac{\delta_1 - \delta_2}{2} \; ,
\ee
and rewrite the S-matrix (\ref{S.DIAG.1}) as
\be \label{S.DIAG.2}
   \tsf{S}_{\mathrm{LP}}  = e^{i\Sigma} \,
  \begin{pmatrix}
  \exp{i \Delta} & 0 \\
  0 & \exp{-i \Delta}   
  \end{pmatrix} 
  \equiv e^{i\Sigma} \exp i \triangle \; .
\ee
This is the polar decomposition of $\tsf{S}_{\mathrm{LP}}$ into a $U(1)$ factor $e^{i\Sigma}$ and an $SU(2)$ matrix $\exp i \triangle$ with unit determinant. To elucidate the physics we use the formalism of \cite{Baym:1990}, which relates the phase exponents, $\delta_i$, to the refractive indices, $n_i$, according to 
\bea \label{PHASE.EXPS}
  \delta_i &=& \omega_{\ell}\, d \, n_i  \equiv \phi_0\, n_i \; \\
  \Sigma &=& \phi_0 \, \frac{n_1 + n_2}{2} \; , \\
  \Delta &=& \phi_0 \, \frac{n_1 - n_2}{2} \; ,
\eea
where $\omega_{\ell}$ is the photon energy, $d$ the duration of propagation through the background and $\phi_{0}$ is the photon phase change without birefringence, i.e. in a classical vacuum. In the weak-field regime, the refractive indices differ only slightly from unity, their vacuum value, such that $\delta n_i \equiv n_i - 1 \ll 1$. We can thus rewrite 
\bea
  \Sigma &=& \phi_0 \, \left( 1 + \frac{\delta n_1 + \delta n_2}{2} 
  \right) \equiv \phi_0 + \sigma \; , \\
  \Delta &=& \phi_0 \, \frac{\delta n_1 - \delta n_2}{2} \; ,
\eea
with $\sigma$ and $\Delta$ being small compared to unity, so that they  may serve as expansion parameters. The overall phase factor, $e^{i\Sigma} = e^{i\phi_0}e^{i\sigma}$, acts as a straightforward multiplication, shifting all incoming polarisations by the common phase, $\Sigma = \phi_0 + \sigma$.

In the generic case, $\delta_1 \ne \delta_2$ or $n_1 \ne n_2$, the second factor, $\exp i \triangle$, in the S-matrix (\ref{S.DIAG.2}) describes birefringence, as each basis polarisation, $\ket{i}$, undergoes its own phase shift, $\pm \Delta$. This may also be inferred by identifying $\tsf{S}_{\mathrm{LP}}$ with the Jones matrix for a birefringent medium \cite{Meschede:2004,Peatross:2025}. For instance, an initial state polarised linearly at an angle of $\beta_0$ from the $1$-axis,
\be \label{IN.LP}
  \ket{\mbox{in}} = \cos \beta_0 \, \ket{1} + \sin \beta_0 \, \ket{2} \; ,
\ee
will turn into the transformed state,
\be 
  \ket{\mbox{in}'} = \tsf{S}_{\mathsf{LP}} \ket{\mbox{in}} =  
  e^{i \Sigma} \left( e^{i \Delta} \cos \beta_0 \, \ket{1} 
  + e^{-i \Delta} \sin \beta_0 \, \ket{2} \right) \; .
\ee
Apart from the overall phase factor, $e^{i \Sigma} $, this is a typical Jones vector for \emph{elliptic} polarisation \cite{Peatross:2025}, with a phase difference of $2 \Delta \equiv \delta_1 - \delta_2$ between the `horizontal' and `vertical' components. The non-flip amplitude measures the `persistence' of the incoming state, hence the overlap
\be \label{M.NF}
  \mathfrak{M}_{\mathrm{nf}} = \braket{\mbox{in}}{\mbox{in}'} = 
  e^{i \Sigma} \left( \cos^2 \beta_0 \,  e^{i \Delta} 
  + \sin^2 \beta_0 \, e^{-i \Delta} \right) \; .
\ee
This result is formally exact, valid to all orders in $\Sigma$ and in particular $\Delta$, as the S-matrix must have the form (\ref{S.DIAG.2}). However, the phases $\delta_i$ will typically be only known through their perturbation series in $\sigma$ and $\Delta$ which in turn will be given in terms of their expansion in the QED coupling $\alpha$. The LO HE 4-photon vertex implies that both $\sigma$ and $\Delta$ are of order $\alpha^2$. For the non-flip amplitude (\ref{M.NF}), the leading order in $\sigma$ and $\Delta$ corresponds to
\be
  \mathfrak{M}_{\mathrm{nf}} \simeq e^{i\phi_0} (1 + i \sigma +
   i \Delta \cos 2 \beta_0) = e^{i\phi_0} (1 + i \sigma )\; , 
\ee
where the last equality is valid for the standard choice $\beta_0 = \pi/4$. In this case, the approximate non-flip probability is thus $P_{\mathrm{nf}} \simeq 1 + \sigma^2$. The fact that this is greater than unity is an artefact of the approximation as the exact result (for $\beta_0 = \pi/4$) is $P_{\mathrm{nf}} = \cos^2 \Delta$. This is a simple example of an apparent unitarity violation that is fixed by `resummation', i.e.\ by working out the exact, all-orders answer. Where this is not possible, higher-order corrections may at least be indicative of the remedy \cite{Heinzl:2021mji,Torgrimsson:2021wcj}.

Of particular interest is the amplitude for polarisation flip into the state
\be 
  \ket{\mbox{f}} = -\sin \beta_0 \, \ket{1} + \cos \beta_0 \, \ket{2} \; ,
\ee 
which is orthogonal to $\ket{\mbox{in}}$. This yields the flip amplitude
\bea 
  \mathfrak{M}_{\mathrm{f}} &=& \braket{\mbox{f}}{\mbox{in}'} = 
  e^{i\Sigma} \left[ \sin\beta_0 \cos \beta_0 \left( e^{i \delta_2} \
  - e^{i \delta_2} \right) \right]
  \nonumber \\
  &\simeq& - e^{i\phi_0} \Delta \sin 2 \beta_0 = - e^{i\phi_0} \Delta   
  \; ,
\eea
with, again, the last identity holding for $\beta_0 = \pi/4$. The exact flip probability is $P_{\mathrm{f}} = \sin^2 \Delta = \Delta^2 + O(\Delta^4)$, such that indeed $P_{\mathrm{nf}} + P_{\mathrm{f}} =1$.
Note that the exact, non-perturbative probabilities are independent of the average~$\sigma$.  In perturbation theory, however, the non-flip and flip amplitudes are governed by both the sum, $\sigma$, and difference, $\Delta$, of the vacuum refractive indices. For low-energy QED this translates into a dependence on the sum and difference of the two LO low-energy couplings in the HE Lagrangian, $\mathcal{L}_{4,0}$ \cite{Schlenvoigt:2016jrd,Karbstein:2022uwf}.

In this paper, however, we are interested in a CP background, to be probed by an LP beam, say in state (\ref{IN.LP}). In this case, the S-matrix is diagonal in the \emph{helicity} basis, $\ket{\lambda} = \ket{\pm}$, with states 
\be
  \ket{\pm} \equiv \frac{1}{\sqrt{2}} (\ket{1} \pm i \ket{2})
\ee
corresponding to right ($+$) and left ($-$) circular polarisation \cite{Baym:1990}. The completeness relation is ${\mathbb{I} = \ket{+}\bra{+} \,+\, \ket{-}\bra{-}}$. Given the LP initial state (\ref{IN.LP}), we need to write the S-matrix in the polarisation basis $\ket{i}$,
\be 
  \tsf{S}_\mathrm{CP} = \sum_{\lambda = \pm} e^{i \delta_\lambda} 
  \ket{\lambda} \bra{\lambda} = \sum_{i,j = 1}^2 \ket{i} 
  \left( \sum_{\lambda = \pm} e^{i \delta_\lambda}  \braket{i}{\lambda} 
  \braket{\lambda}{j} \right) \bra{j} \; .
\ee
The expression in parenthesis defines the entries of the S-matrix we are looking for. They take on the explicit form
\bea \label{S.CP}
  \tsf{S}_\mathrm{CP} &=& U \begin{pmatrix}
  e^{i \delta_+} & 0 \\ 0 & e^{i \delta_-}
  \end{pmatrix} U^\dagger  = 
  e^{i \Sigma} \begin{pmatrix}
  \phantom{-}\cos \Delta & \sin \Delta \\
  -\sin \Delta & \cos \Delta 
  \end{pmatrix} \nn \\
  &\equiv& e^{i \Sigma} R(\Delta) \; ,
\eea
with $\Sigma \equiv (\delta_+ + \delta_-)/2$ and $\Delta \equiv (\delta_+ - \delta_-)/2$. The S-matrix (\ref{S.CP}) is analogous to (\ref{S.DIAG.2}), though no longer diagonal. Instead we find a polar representation, given by the product of a global $U(1)$ phase shift, $e^{i \Sigma}$ and a \emph{rotation}, $R(\Delta) \in SO(2)$, by an angle $\Delta$. (We use the same symbols as for LP to avoid cluttered notation.) The first equality in (\ref{S.CP}) is nothing but the statement that $\tsf{S}_\mathrm{CP} $ is diagonalised by the unitary matrix
\be \label{U.LP.CP}
  U = \frac{1}{\sqrt{2}} \begin{pmatrix}
  1 & 1 \\
  i & -i
  \end{pmatrix} \; ,
\ee 
which transforms the linear polarisation basis $\ket{i}$ to the circular polarisation or helicity basis $\ket{\lambda}$. (Recall that the columns of (\ref{U.LP.CP}) are the images of the basis vectors.) To study the fate of the LP probe state (\ref{IN.LP}), we act with the CP S-matrix (\ref{S.CP}) and find
\be
  \ket{\mbox{in}'} = \tsf{S}_{\mathsf{CP}} \ket{\mbox{in}} = 
  e^{i\Sigma} \left[ \cos(\beta_0 - \Delta) \, \ket{1} 
  + \sin(\beta_0 - \Delta) \, \ket{2} \right] \; .
\ee
This state corresponds to photons polarised linearly at angle $\beta_0 - \Delta$. The initial polarisation has thus been rotated by an amount of $\Delta$. In other words, the background acts like an optically active \emph{chiral} medium similar to a solution of dextrose or a chiral metamaterial \cite{Wang:2016}. As the background has preferred handedness (chirality),  different helicities (positive or negative) acquire different phases upon traversing the background (or chiral medium), 
\be
  \tsf{S}_{\mathsf{CP}} \ket{\pm} = e^{i\delta_\pm} \ket{\pm} \; ,
\ee
as is obvious from the diagonal form (\ref{S.CP}) in the helicity basis.
For incoming linear polarisation, the non-flip and flip amplitudes become
\bea 
  \mathfrak{M}_\mathrm{nf} &=& \braket{\mbox{in}}{\mbox{in}'} =
  e^{i\Sigma} \, \cos \Delta \simeq e^{i\phi_0} \, (1 + i \sigma)
   \; , \\
  \mathfrak{M}_\mathrm{f} &=& \braket{\mbox{f}}{\mbox{in}'} =
  - e^{i\Sigma} \, \sin \Delta \simeq - \Delta \,  e^{i\phi_0} \; .
\eea
Interestingly, the exact non-flip and flip probabilities are the same as in the LP case, given by $\cos^2 \Delta$ and $\sin^2 \Delta$, respectively. Note that the amplitudes are independent of the initial rotation angle, $\beta_0$, which implies that non-flip and flip amplitudes are simply given by the following matrix elements (in the LP basis),
\bea
  \mathfrak{M}_\mathrm{nf} &=& \bra{1} \tsf{S}_\mathrm{CP} \ket{1} 
  \; , \label{M.NF.12} \\
  \mathfrak{M}_\mathrm{f} &=& \bra{2} \tsf{S}_\mathrm{CP} \ket{1}  
  \label{M.F.12}\; ,
\eea
irrespective of the initial direction of (linear) polarisation.

The different combinations of incoming (= probe), background and outgoing polarisations are summarised in \tabref{tab:birefs}. 
\begin{table}[h!!]
    \centering
    \begin{tabular}{ r | c | c | c | }
    & Input (Probe) & Background & Output\\ 
    \hline
   (i) &    LP  &  LP &  EP \\
   (ii) &   LP & CP &  LP'  \\
   (iii) &   CP & LP &  EP \\ 
   (iv) &   CP & CP & CP
\end{tabular}
\caption{The effects on an incoming probe beam for varying polarisations (LP = linearly polarised, CP = circularly polarised, EP = elliptically polarised). All outputs are phase shifted compared to the input. Circular birefringence, which is the focus of our discussion, corresponds to case (ii) with LP' denoting a rotation of the input LP.}\label{tab:birefs}
\end{table}

Often, the effect of vacuum polarisation on a probe laser \emph{beam} is studied, which corresponds to replacing in states of photons with a classical field comprising positive and negative frequencies (see e.g. \eqnref{eqn:bg1}), i.e. studying the `$0 \to 0$' process. The connection can be made with the above analysis by adding the out state from a photon of fixed polarisation and frequency to its complex conjugate. The results in the `Output' column of \tabref{tab:birefs} can then be understand in terms of classical beams. For example an initially CP probe beam of the form \eqnref{eqn:bg1}, becomes elliptically polarised when propagating through an LP background, which can be understood as a \emph{rotation} of the initial CP beam into another CP beam with the opposite chirality, by angle $\Delta$, with each beam acquiring an overall phase shift of $\Sigma$.
\begin{figure}[h!]
\includegraphics[scale=0.2]{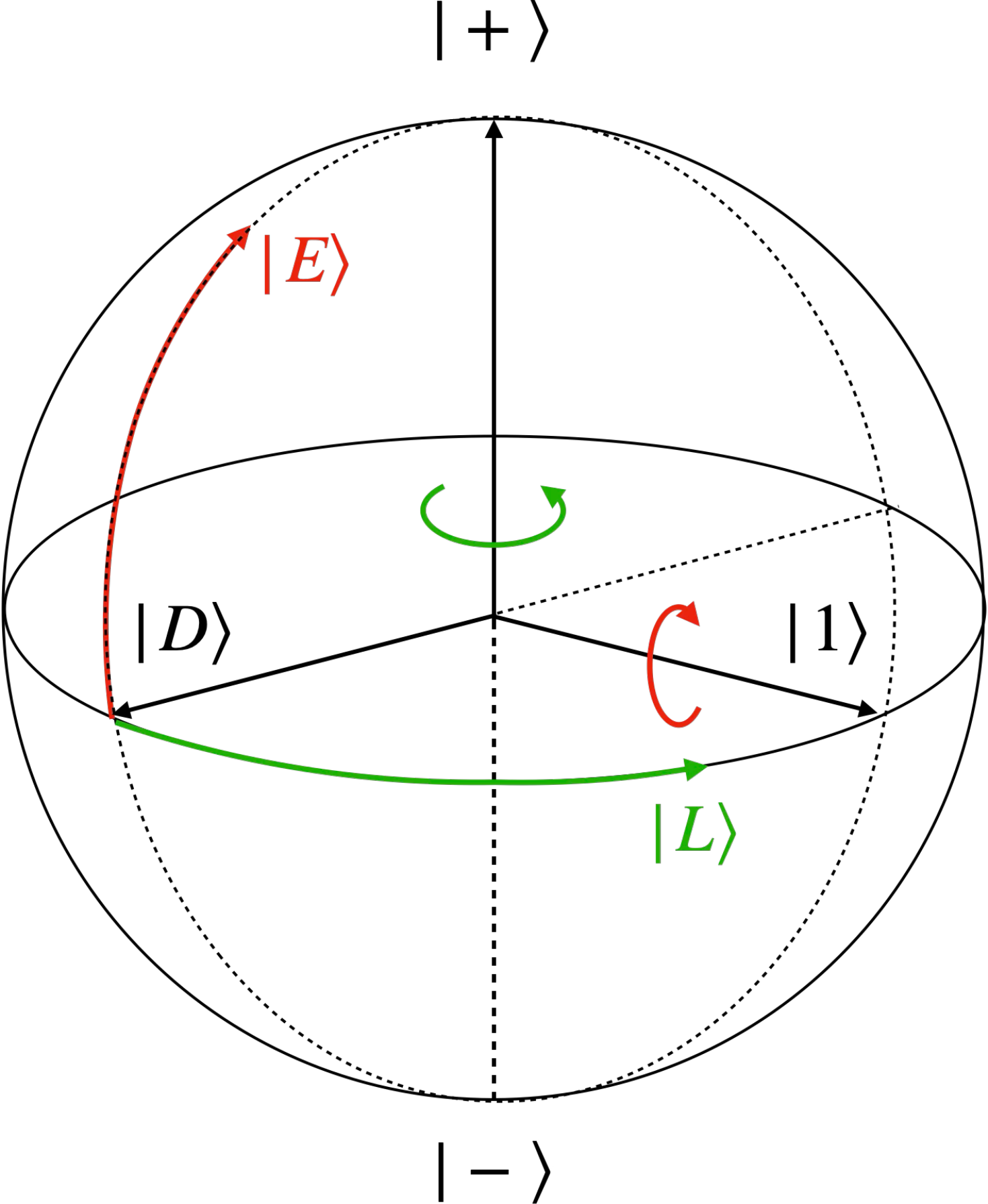}
\caption{\label{fig:PS1} Poincar\'e sphere with polarisation changes for linear and circular birefringence [cases (i) and (ii) of Table~\ref{tab:birefs}]. The input state is $\ket{D} = (1/\sqrt{2}) (\ket{1} - \ket{2})$. For linear birefringence, this is mapped to an elliptically polarised state $\ket{E}$ (red arrows), for circular polarisation (a chiral background), $\ket{D}$ is mapped to a rotated LP state, $\ket{L}$ (green arrows).}
\end{figure} 

\begin{figure}[h!]
\includegraphics[scale=0.2]{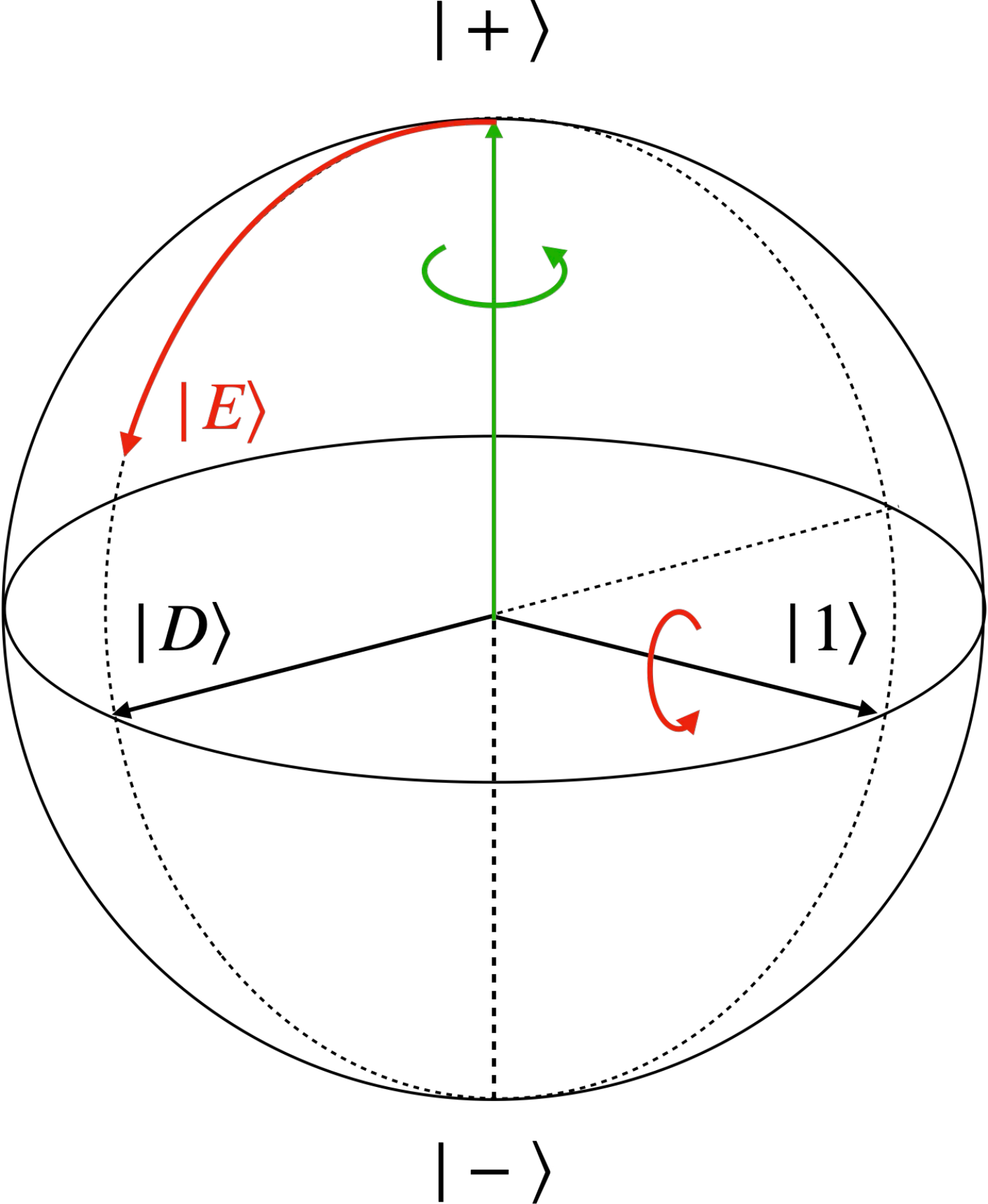}
\caption{\label{fig:PS2} Poincar\'e sphere with polarisation changes for a CP probe [cases (iii) and (iv) of Table~\ref{tab:birefs}. The input state is $\ket{+}$ which is a fixed point, hence mapped to itself, for a CP background (green arrows).  For an LP background, $\ket{+}$ is mapped to an elliptically polarised state $\ket{E}$ (red arrows), unless the phase difference is fine tuned to $\pi/2$ (as for a quarter waveplate) for which one reaches the equator, hence an LP configuration.}
\end{figure}

A particularly intuitive graphical overview is given by the Poincar\'e sphere \cite[Ch.~XII]{Poincare:1892}\footnote{Interestingly, Poincar\'e entitles this chapter as `Circular polarisation -- theory of M.~Mallard' which presumably refers to the second volume of Mallard's treatise on crystallography \cite{Mallard:1884}. While the latter discusses birefringence at length, we could not find any precursor of the Poincar\'e sphere there.} and \cite{Ives:2004,Zangwill:2012,Dennis:2017},  where a generic point represents elliptic polarisation, the equator collects all possible linear polarisations while north and south poles correspond to right, $\ket{+}$, and left, $\ket{-}$, circular polarisation, respectively. Orthogonal vectors are given by antipodal points. Any change in polarisation caused by a background field (and described by the corresponding S-matrix) is represented by a great circle segment on the sphere which results from a rotation around the `fast axis' of background, hence the eigenvector associated with the smaller phase shift, $\min(\delta_1, \delta_2)$ or $\min(\delta_+, \delta_-)$, as the velocities of light are $v_i = 1/n_i \simeq 1 - \delta_i$ and analogously for $v_\pm$. Cases (i) and (ii) are depicted in Fig.~\ref{fig:PS1}, cases (iii) and (iv) in Fig.~\ref{fig:PS2}.

\section{Vacuum circular birefringence in a plane wave background}

\begin{figure}[h!!]
\centering
\includegraphics[width=6cm]{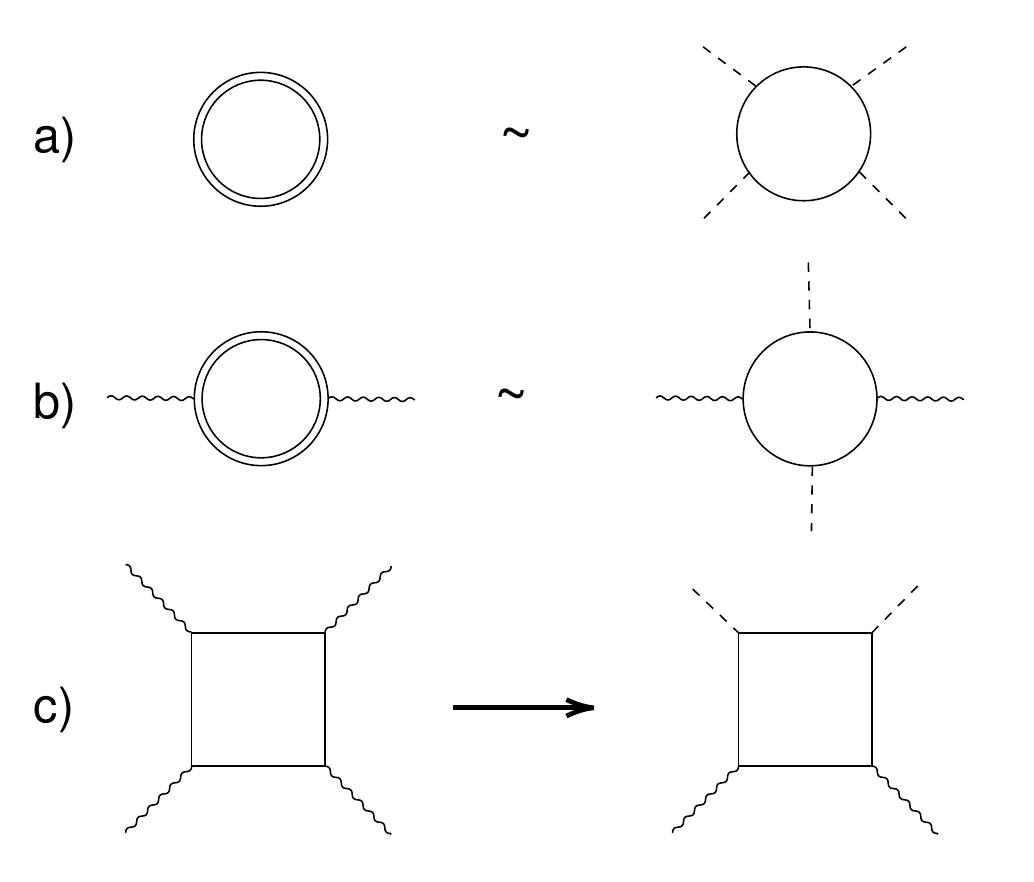}
\caption{a) the LO term in field strength of the HE Lagrangian ($0 \to 0$ process, i.e. a dressed vacuum loop), with the probe and background as classical fields (dashed lines); b) the LO expansion in field strength of the $1\to 1$ process of photon scattering in a plane-wave background (polarisation tensor); c) $2\to 2$ photon-photon scattering with one incoming and one outgoing photon (wavy lines) replaced by a classical plane wave background field.} \label{fig:approaches1}
\end{figure}

In the remainder of the paper we focus on the calculation of the polarisation flip and non-flip amplitudes for a CP background, hence the matrix elements of $\tsf{S}_\mathrm{CP}$, in QED and its low energy approximation. We will compare three approaches based on: (i) the derivative expansion of the LCFA; (ii) the derivative corrections (\ref{L.4.2}) to the HE Lagrangian, and (iii) the low-energy expansion of the QED amplitude for $2\to 2$ scattering. All relevant amplitudes are depicted in Fig.~\ref{fig:approaches1} in terms of Feynman diagrams stemming from $0 \to 0$, $1 \to 1$ and $2 \to 2$ photon processes, respectively. 

\subsection{Derivative corrections to the LCFA}

The main advantage of plane wave backgrounds is their high symmetry leading to super-integrable dynamics \cite{Heinzl:2017zsr}. As a result, one can derive exact results, valid to all orders in field strength and/or energy. Alternatively, one can view these results as being due to a resummation of the weak-field or low-energy expansions.  Plane-wave backgrounds thus provide a useful test bed for assessing the quality of the derivative expansion described in the previous section. 

To proceed, we use (\ref{M.F.12}) and express the probability (\ref{P.FLIP}) as 
\be
  \tsf{P}_{21} = |\bra{2} \tsf{S}_\mathrm{CP} \ket{1}|^2 
  = \bigg| \eps_{2}^{\ast\,\mu} \, 
  \mathfrak{M}^{(21)}_{\mu\nu} \, \eps_{1}^{\nu} \bigg|^{2} \; ,
\ee
where we have mapped the states $\ket{1}$ and $\bra{2}$ back to the polarisation 4-vectors $\eps_1$ and $\eps_2^*$ for incoming and outgoing photons, respectively. The tensor amplitude $\mathfrak{M}^{(21)}$ represents an integral over the background phase,
\be \label{M.21.INT}
  \mathfrak{M}^{(21)} = \int \mathfrak{M}^{(21)}
  (\vphi) \, d\vphi \; ,
\ee
suppressing Lorentz indices for convenience. The integrand, $\mathfrak{M}^{(21)} (\vphi) $, has been found in e.g. \cite{narozhny69,baier75a,Meuren:2013oya,Dinu:2013gaa}. Below, we use the double-integral representation given in \cite{King:2023eeo}:
\bea \label{M.21.INT.REP}
  \mathfrak{M}^{(21)}_{\mu\nu}(\vphi) &=& \frac{\alpha}{8\pi\eta}
  \int_{0}^{\infty}\frac{d\theta}{\theta} \int_{0}^{1}dr  \exp\left[\frac{i
  \theta\mu(\theta)}{2\eta r(1-r)}\right]  \nonumber \\
  && \left\{\frac{2(1-2r)^{2}}{r(1-r)}\left(\bar{a}_{\mu}-\av{a_{\mu}}
  \right)\left(a_{\nu}-\av{a_{\nu}}\right) \right.\nn \\
  && \left.\!\! -\frac{2}{r(1-r)}\left(a_{\mu}-\av{a_{\mu}}\right) 
  \left(\bar{a}_{\nu}-\av{a_{\nu}}\right)\right\}\!\!.
\eea
Here $\eta=\vkap \cdot \ell / m^{2}$  is the photon energy parameter, while the integration variable $r$ denotes a light-front momentum fraction of the virtual positron or electron. We have abbreviated $a = a(\vphi+\theta/2)$, $\bar{a}=a(\vphi-\theta/2)$ and introduced the phase-window average
\be
  \av{f}=\theta^{-1}\int^{\vphi+\theta/2}_{\vphi-\theta/2} f(x) dx
\ee  
such that the Kibble mass squared, $\mu(\theta)=1+\av{a}\cdot\av{a} - \av{a\cdot a}$ (in units of $m^{2}$). The result (\ref{M.21.INT.REP}) is valid for any centre-of-mass energy and field intensity, but specific to a plane wave background. Hence, (\ref{M.21.INT.REP}) is an example of an exact result, valid to all orders in $\eta$ and the background intensity parameter $\xi$ defined in (\ref{eqn:bg1}).

As is well-known \cite{narozhny69,baier75a,Dinu:2013gaa}, a comparison can be made between the plane wave result and the HE approach by calculating the LCFA of the former and taking the limit of small strong-field parameter, $\chi = \xi \eta \ll 1$. (The small-$\chi$ limit is consistent with a low-energy limit, since the LCFA requires $\xi \gg 1$, which needs to be over-compensated by small $\eta \ll 1$.) However, as was noted in \cite{King:2023eeo}, the standard LCFA result is highly suppressed for the case at hand and does not reproduce the correct small $\chi$ behaviour. Instead, one must include derivative corrections to the LCFA itself. (This is similar to the `LCFA+' extensions of tree-level processes \cite{DiPiazza:2018bfu,Ilderton:2018nws,King:2019igt}, but here without the need for filters, and extendable to arbitrarily high orders of derivatives.) We assume that the background wave has many cycles so that $\psi'(\vphi)/\psi(\vphi) \ll 1$, whence derivatives of the envelope can be ignored, and use the CP potential from \eqref{eqn:bg1}. Defining $\xi(\vphi) = \xi \psi(\vphi)$, substituting
\be
  \theta \to 2u\left[\frac{r(1-r)\eta}{\xi^{2}(\vphi)}\right]^{1/3},
\ee
in (\ref{M.21.INT.REP}) and expanding in the small parameter, $1/\xi$, yields an Airy-type kernel in $u$:
\bea
  \mathfrak{M}^{(21)}_{\mu\nu}(\vphi) &=& \frac{\alpha}{8\pi\eta} 
  \int_{0}^{1}dr \int_{0}^{\infty}du\, \left(1+i k_{5} u^{5}+\cdots\right)    
  \nonumber \\
  && \; \times \sum_{j\ge 1} c_{j} u^{j}\,\exp\left[i\left(z u +
  \frac{u^{3}}{3}\right)\right] ,
\eea 
where $z = [\chi(\vphi) r(1-r)]^{-2/3}$ with $\chi(\vphi)=\chi \psi(\vphi)$, $\chi=\eta \xi$ and ${k_{5} = 4/[45z\xi^{2}(\vphi)]}$. The $c_j$ are phase dependent expansion coefficients (see below). The $u$-integral can be performed by replacing powers of $u$ in the pre-exponent with derivatives:
\bea
  \mathfrak{M}^{(21)}_{\mu\nu}(\vphi) &=& \frac{\alpha}{8\eta} 
  \int_{0}^{1} dr \left(1+ k_{5} \frac{d^{5}}{dz^{5}}+\cdots\right)    
  \nonumber \\
  && \times \sum_{j \ge 1} (-i)^{j}c_{j} \frac{d^{j}}{dz^{j}} \left[\Ai(z) + i 
  \Gi(z)\right], \label{eqn:LCFA2}
\eea 
where $\Gi$ is the first Scorer function \cite{olver97}. Expression \eqref{eqn:LCFA2} manifestly exhibits the derivative expansion of the LCFA: each higher derivative with respect to $z$ is associated with a higher power of the energy parameter $\eta$. Depending on the parameter regime of interest and the relative size of $\chi$ and $\eta$, we could choose to expand the Airy-type functions in $\chi$ or use the derivative expansion to include higher powers in $\eta$. The LO term in the derivative expansion \eqref{eqn:LCFA2} is the $c_{1}$ term, but for the CP background we find that $c_{1} \propto \sin (2\vphi)$, which is rapidly oscillating and integrates to zero in the long-pulse limit. Thus the NLO term from $c_{2}$ becomes dominant, and we see the same disruption of hierarchy in derivative terms as in the effective field theory approach. For now, we concentrate on birefringence at small $\chi$ and use the asymptotic result \cite{nist_dlmf}:
\be
  \Gi(z) \sim \frac{1}{\pi z}\sum_{k=0}^{\infty} 
  \frac{(3k)!}{k!(3z^{3})^{k}} .
\ee
(We neglect the contribution from $\Ai(z)$, which is exponentially suppressed for small $\eta$ i.e.  large $z$). This allows the probability to be expanded in the form:
\bea 
\tsf{P} = \bigg|\alpha \sum_{i,j=0}^{\infty} C_{i,j}\,\chi^{2i}\eta^{2j}\bigg|^2
\eea
where $C_{i,j}$ are real coefficients. We eventually find, expanding up to $O(\eta^{6})$:
\be
  \tsf{P} = \frac{\alpha^{2}}{\pi^{2}}\bigg|\frac{2\chi^{2}}{315}\left(1+
  \frac{8\eta^{2}}{55}+\frac{24\eta^{4}}{1001}\right) \mathcal{I}_{2}  + 
  \frac{4\chi^{4}}{693} \mathcal{I}_{4} + \frac{64 \chi^{6}}{3861} \mathcal{I}_{6} 
  \bigg|^{2},  \label{eqn:CPflipProb1}
\ee
where\footnote{The term $\chi^{4}\eta^{2}$ evaluates to zero when the contribution of $k_{5}$ is taken into account.} $\mathcal{I}_{2n} = \int \psi^{2n}(\vphi)d\vphi$, independent of $\eta$. We note the first term of the amplitude in \eqnref{eqn:CPflipProb1}, which scales with  $\sim O(\eta^{0}\chi^{2})$, is a factor $\eta$ larger than the LO of the result, for the `standard' vacuum linear birefringence case of photon helicity flipping (CP) in an LP background, which scales as $\sim O(\chi^{2}/\eta)$ (see e.g. \cite{Dinu:2013gaa}). The LO term in \eqnref{eqn:CPflipProb1} corresponds to the NLO order in the derivative expansion. For the plane-wave set-up of this section, the standard Mandelstam variables become
\be
s = (\vkap+\ell)^{2} = 2 m^{2} \eta; \quad t=-s; \quad u=0. \label{eqn:mandelstam1}
\ee
Therefore, the NLO term of the LCFA having one higher power of $\eta$  corresponds to an amplitude with one higher power of the Mandelstam variables, which is exactly what was found for the NLO terms (\ref{M.NLO}) in the derivative expansion in the previous section. 

To confirm our findings, we recalculate the flip amplitude from the effective field theory of Section~II.

\subsection{Derivative corrections to the HE Lagrangian}

We use the low energy limit of the HE Lagrangian together with its NLO derivative correction given by \eqref{HE.LO} and \eqref{L.4.2}.  Following \cite{Davila:2013wba} we express the effective Lagrangian in terms of the total electromagnetic field tensor, which for LP photons flipping polarisation in a CP plane wave background is written $F^{\mu\nu} = \mathcal{F}^{\mu \nu} + f^{\mu \nu} + f'^{\mu \nu} $, where $\mathcal{F}$ is the plane wave background field with the vector potential given in \eqref{eqn:bg1}, while $f$ and $f'$ correspond to the incoming and outgoing LP photons,
 \begin{eqnarray}
     f^{\mu \nu} &=& \frac{1 }{ \sqrt{ 2 V \ell ^0}} e^ { i \ell \cdot x}
     \left(\ell^\mu \eps^\nu_{1,2} - \ell^\nu \eps^\mu_{1,2} \right) \\
     f'^{\mu \nu} &=& \frac{1 }{ \sqrt{ 2 V \ell^{\prime\,0}}} e^ {- i \ell' \cdot x}
     \left(\ell'^\mu \eps'^{\nu \, *}_{1,2} - \ell'^\nu \eps'^{\mu \, *}_{1,2} 
     \right) \; .
\end{eqnarray}
$V$ is a volume factor, $\ell$ and $\ell'$ are the momenta of the incoming and outgoing photons and $\eps$, $\eps'^*$ their polarisation vectors. The terms contributing to the photon polarisation flip involve the contraction of $f$ and $f'$, but must also include the mixed contraction $a^{(+)} \cdot a^{(-)}$, as any self-contraction of $a^{(+)}$ or $a^{(-)}$ will oscillate on the fast time scale of the plane wave and integrate to zero. This is the mathematical explanation for  why the LO terms from the Lagrangian are vanishing, as they typically involve $(a^{(\pm)})^2$ terms which average to zero when integrated over space-time. We use first-order perturbation theory to calculate the probability, 
\be
  \tsf{P} = V\int \frac{d^{3}\ell}{(2\pi)^{3}} \bigg|-i\int d^{4}x\,
  \mathcal{L}_{\tsf{eff}}\bigg |^{2} \; ,
\ee
and make the required substitutions in the Lagrangian to find
\begin{eqnarray}
   \mathcal{L}_{\tsf{eff}} &=& i\frac{8\alpha ^{2}}{315m^{6}}\frac{e^{i x
   \cdot \delta \ell}}{2V\ell ^{0}}(\ell \cdot \varkappa )^{2}\left\{2(\ell 
   \cdot \varkappa ) \right. \nn \\    
   && \left. +5(e_{1} \cdot \ell )(e_{2} \cdot \delta \ell )-\frac{\delta 
   \ell^{0}}{\ell^{0}} (\ell \cdot \varkappa)\right\}+O\left(\delta \ell^{2}
   \right) \; , \nn \\
\end{eqnarray}
where $\delta \ell = \ell' - \ell$. We have used the photon polarisation states from \eqref{eqn:photonPol1} and expanded the result in powers of $\delta \ell$. When the spacetime integral is performed, a delta function in $\delta \ell$ is clearly formed, which will be evaluated when the scattered momentum is integrated over. We then obtain the same probability for flipping as the first term in \eqref{eqn:CPflipProb1}.
The NNLO of the derivative expansion, given by $\mathcal{L}_{4,4}$ in \eqref{L.4.4} can also be included in this analysis; we find for the scenario studied here that it yields terms that are again proportional to a fast oscillation term as for the LO term; therefore $\mathcal{L}_{4,4}$ does not contribute. Indeed this is also consistent with the derivative LCFA analysis as $\mathcal{L}_{4,4}$ would contribute terms of order $\sim \eta\chi^{2}$, which were integrated over and thus are absent in \eqref{eqn:CPflipProb1}.

Throughout, we have assumed derivatives of the pulse envelope are zero. If we relax this assumptions, we can estimate their contribution. The function $\psi(\varphi)$ in the vector potential is a function of $\varphi/\Phi$, where $\Phi$ represents the characteristic phase duration of the pulse. Every power $\mathcal{F}^{\mu \nu}$ of the background then contains a term proportional to the derivative $(1/\Phi) \psi'(\varphi/\Phi)$. The elastic LO contribution from $\mathcal{L}_{4,0}$ is
\begin{equation}
\begin{split}
    \mathcal{L}_{4,0} &= \frac{4 \alpha ^2 m^2\chi^2  e^{i(\ell - \ell ')
    \cdot x}}{15 V \sqrt{\ell ^0 \ell '^0}} \left[ \psi 
    \left( \frac{\varphi}{\Phi} \right)  \psi' \left( \frac{\varphi}{\Phi} \right)
    \cos( 2\varphi) \frac{1}{\Phi}  \right.\\
    & \left. + \left(\psi'^2 \left( \frac{\varphi}{\Phi} \right)  
    \frac{1}{\Phi ^2}  - \psi^2 \left( \frac{\varphi}{\Phi} \right)  \right) 
    \cos (\varphi) \sin (\varphi)  \right] \, .
\end{split} \label{eqn:LI40a}
\end{equation}
If the pulse profile is symmetric around the origin $\psi(\vphi/\Phi) = \psi(-\vphi/\Phi)$, the integral of \eqref{eqn:LI40a} over $\vphi$ gives exactly zero (using $\ell=\ell'$), despite including the pulse duration. The NLO terms from $\mathcal{L}_{4,2}$, however, yield  a non-zero contribution,
\bea
    \mathcal{L}_{4,2} &=& \,\frac{-4 i \alpha  \chi^2 (\ell \cdot \kappa) 
    e^{i(\ell - \ell ')\cdot x}}{315 \pi V \sqrt{\ell ^0 \ell '^0}} \nn  \\ 
    &\times& \left\{ \psi^2\left(\frac{\varphi}{\Phi}\right)  +
    \frac{1}{\Phi ^2}\left[\psi'^2\left(\frac{\varphi}{\Phi}\right) 
    - \psi\left(\frac{\varphi}{\Phi}\right) \psi''\left(\frac{\varphi}{\Phi}
    \right)\right]\right\} \, .  \nn \\
\eea
Therefore, derivative corrections in the probability will appear at $\sim O(\Phi^{-2})$. (For a laser pulse, typically $\Phi \sim 2\pi N$ where $N \geq 1$ is the number of cycles.)

\subsection{Photon-photon scattering}

The general result for $2\to2$ photon-photon scattering can be directly related to $1\to1$ scattering in a classical plane wave by a replacement prescription.  It can be shown \cite{Heinzl:2024cia} that the probability can be written as
\bea
  \tsf{P} = V\int \frac{d^{3}\ell'}{(2\pi)^{3}} \big|T_{\tsf{fi}}\big|^{2}; 
  \quad T_{\tsf{fi}} = K(\ell'-\ell) \frac{2V}{\vkap^{0}} 
  \left[\mathfrak{M}^{2\to2}\right]_{1\to1} \; , \nn \\
\eea
where $K$ is a \emph{form factor}. For a weakly-focussed beam it can be expressed as a Fourier integral over the intensity distribution,
\bea 
  K(q) = (m\xi \vkap^{0})^{2} \int \psi^{2}(\vphi) \mbox{e}^{iq\cdot x}
  \, d^{4}x \; .
\eea
The labels on the invariant amplitude $\mathfrak{M}$ correspond to replacing the momentum and polarisation of one incoming and one outgoing photon with the momentum and polarisation of a Fourier mode of the plane-wave background. For the CP background \eqref{eqn:bg1} one obtains
\bea
  \tsf{P} = \bigg|
  \frac{1}{\eta}
  \left[\bar{\mathfrak{M}}^{2\to2}\right]_{1\to1} \int \frac{1}{2}\xi^{2}
  \psi^{2}(\vphi) d\vphi \bigg|^{2} \; , \label{eqn:PDT}
\eea
where the bar on $\mathfrak{M}$ indicates that the wave function normalisation factors of the form $(2Vk^{0})^{-1/2}$ have been set to unity. We can compare with literature results for helicity/polarisation amplitudes through the definition
\be
  \mathfrak{M}^{2\to2}=\eps^{i}_{k_{1}}\eps^{j}_{k_{2}}\eps^{i'}_{k_{3}}
  \eps^{j'}_{k_{4}}\mathfrak{M}^{2\to2}_{iji'j'} \; ,
\ee
with polarisation vectors $\eps_k$ and a summation over polarisation labels  $i,j,i',j' = 1,2$ implied. We then perform the replacement in \eqref{eqn:PDT}, employing
\be
  \mathfrak{M}^{2\to2}_{1+-2} = \frac{i}{2} \left(\mathfrak{M}_{1212}-
  \mathfrak{M}_{1122}\right).
\ee
Using the low-energy expansion of the helicity amplitudes available in the literature \cite{DeTollis:1964una,DeTollis:1965vna,Colwell:2015wna}, we infer that  the LO terms are given by the difference of the first two equations in (\ref{M.LO}) and thus are proportional to $s^{2}-t^{2}$. In a plane-wave background $t=-s$, which is why the LO contributions cancel. (For a general background, the scattered and incident photon momentum may differ, and the leading order may no longer vanish.) For the plane wave case, the probability depends on linear combinations of the NLO terms listed in (\ref{M.NLO}). Specifically, we find a proportionality $\propto s^{3}-t^{3}$, which no longer vanishes when $t = -s$. Using the generating functions in \cite{DeTollis:1965vna}, one can calculate arbitrary high powers in the low-energy expansion with the result
\be
  \big|\bar{\mathfrak{M}}^{2\to2}_{1+-2}\big|^2 = \bigg|\frac{4\alpha
  \eta^{3}}{315\pi}\left(1+\frac{8 \eta^{2}}{55} + 
  \frac{24 \eta^{4}}{1001}-\frac{6272 \eta^{6}}{415701}\right)\bigg|^2.
\ee
Putting everything together, the probability then agrees with (and goes beyond) all terms in \eqref{eqn:CPflipProb1} proportional to the field squared. To NLO, we therefore find the three methods discussed in this section to be equivalent.

The evaluation of the QED amplitude for the box diagram of photon-photon scattering suggests that if circular birefringence is calculated in a set-up where $\ell' \neq \ell$, for example if the background is not a plane wave, then the LO contribution will not necessarily disappear, although it may still be strongly suppressed. In the next section, we look at the example of a rotation standing wave background.

\section{Rotating standing wave background}

\begin{figure}[h!!]
\centering
\includegraphics[width=8cm]{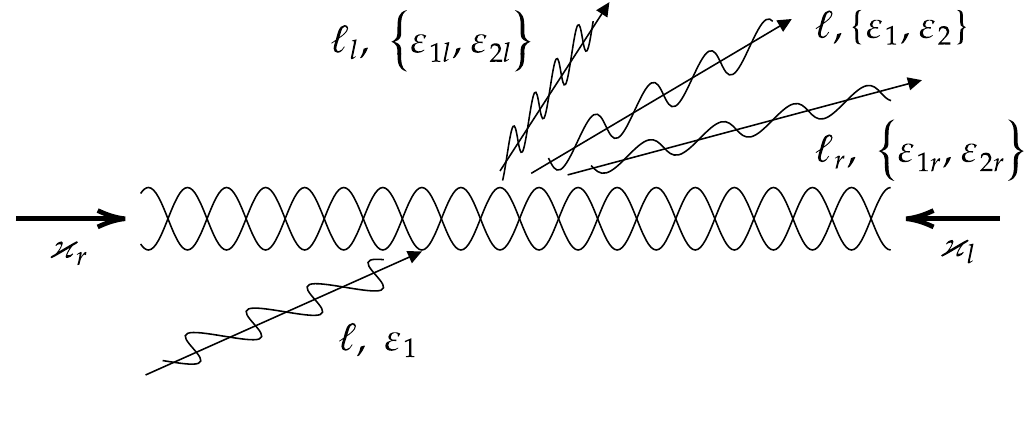}
\caption{Sketch of the circular standing-wave scenario: The \emph{linear} polarisation of probe photons flips due to scattering off the standing CP plane-wave background (with momentum and polarisation labels). The main signal has unchanged momentum $\ell$ (forward scattering), while the photons on the side peaks have altered momenta $\ell_{r,l}$ and polarisations $\neq \varepsilon_{1,2}$.} \label{fig:sketch_standing_wave}
\end{figure}

In a Volkov approach, the electromagnetic background must be close to a plane wave for the calculation to yield a good approximation. Employing the derivative expansion of the QED effective action, however, allows for treating backgrounds that are not restricted to being a plane wave. As an example of this kind we consider a standing-wave background formed by two counter-propagating CP plane waves. We choose a particular configuration where the plane waves collide head on, with opposite helicities and such that the electric fields of each plane wave are parallel (and likewise for the magnetic fields). We call this the `rotating standing wave' set-up. The corresponding vector potential takes the form:
\begin{equation}
    A = a^{(+)}(\varphi_{r})  + a^{(-)}(\varphi_{r}) +  a^{(+)}(\varphi_l)  + 
    a^{(-)}(\varphi_l)
\end{equation}
where $\varphi_{r} = \varkappa_r \cdot x = \vkap_{r}^{+}x_{+}$ and $\varphi_l = \varkappa_l \cdot x = \vkap_{l}^{-}x_{-}$. Again, we consider the incoming photon to be in linear polarisation state $\ket{1}$ corresponding to polarisation 4-vector $\eps_{1}$. However, as the background now differs from a plane wave, the scattered photon momentum $\ell'$ is not automatically equal to the incoming momentum $\ell$. Instead, photon scattering in the standing wave can be understood as a collision of three plane waves, the kinematics of which have been well-studied in the literature \cite{lundstroem_PRL_06,Gies:2017ezf,Aboushelbaya:2019ncg,macleod2024fundamental}. One finds the photon may be scattered into three peaks: a central peak with momentum $\ell$ and two neighbouring peaks with momentum $\ell \pm \Delta \ell$ where $\Delta \ell = \vkap_{r}-\vkap_{l}$. Because the photon polarisation is transverse to the momentum, this also means there are three different polarisation bases, $\{\eps_{1},\eps_{2}\}$, $\{\eps_{1l},\eps_{2l}\}$, $\{\eps_{1r}, \eps_{2r}\}$: the Hilbert space of photon polarisation is no longer two dimensional. The set-up is depicted in \figref{fig:sketch_standing_wave}. The signature of the NLO terms can be identified by considering the polarisation-flipped signal in the central peak, which we find is:
\begin{eqnarray}
     \mathfrak{M}^{(21)}&\approx&\frac{4\alpha}{315\pi} \frac{1}{2  V
     \sqrt{l^0 l'^0}} \nonumber \\
     &\times& \int \drm ^4 x \left[ (\ell \cdot \varkappa_r) \, 
     \chi_{r}^2(\varphi_{r}) + (\ell \cdot \varkappa_l)  \, 
     \chi_{l}^2(\varphi_l)\right]  e^{i \delta \ell\cdot x},\nn \\
\end{eqnarray}
where $\delta \ell = \ell' - \ell$. We have already set $\ell'=\ell$ in the pre-exponent in anticipation of the delta function that will be generated and integrated over. In addition, we have only included terms to zeroth order in $\varkappa_r \cdot \varkappa_l$, assuming that  $\varkappa_r \cdot \varkappa_l \ll \ell \cdot \varkappa_{r}$ and $\varkappa_r \cdot \varkappa_l \ll \ell \cdot \varkappa_{l}$. In this limit, the polarisation flip amplitude is simply the sum of contributions from each of the two plane waves that comprise the standing wave background. However, at the probability level, an interference term is generated, which is absent from the plane wave treatment. For the probability, we thus find (after mod-squaring and integrating over momentum),
\begin{eqnarray}
    \tsf{P} &=&\bigg| \frac{2 \alpha}{315 \pi} \int \drm \varphi\,\left[ 
    \chi_{r}^2(\varphi) +\chi_{l}^2(\varphi)\right] \bigg|^{2}, \label{eqn:PCPsw1}
\end{eqnarray}
where $\chi_{r,l}=\chi_{r,l}(\vkap_{r,l}\cdot x)$. We note that if $\eta_{r} \gg \eta_{\ell}$ or vice versa, the result tends to the plane-wave result (e.g. \eqnref{eqn:CPflipProb1}) with a wave vector originating from the larger of $\eta_{r}$ or $\eta_{l}$. We can also rephrase the result via the derivative-modified LCFA by calculating $\chi$ for the full background and dropping fast oscillating factors (which will give only a negligible contribution once the phases are integrated over), leading to $\chi^{2} = \chi_{r}^{2} +\chi_{l}^{2}$. Hence, for the \emph{central peak} of the standing-wave background, assuming that $\vkap_{l}\cdot \vkap_{r} \ll \vkap_{l}\cdot \ell$ and $\vkap_{l}\cdot \vkap_{r} \ll \vkap_{r}\cdot \ell$, the result \eqref{eqn:PCPsw1} can be recovered by simply using the plane-wave result \eqref{eqn:CPflipProb1} and calculating $\chi$ for the standing-wave background. 

For the sidepeaks (an effect beyond a plane wave background, which the LCFA cannot replicate), the main contribution to the polarisation-flip probability is from the regular, LO terms of the HE Lagrangian. This, however, is not the dominant polarisation-flip signal. Instead, the LO terms are proportional to the momentum exchange $\delta \ell$; if the probe energy is much larger than the central frequency of the standing wave, then $\delta \ell \ll \ell$. Specifically, due to momentum conservation, $\ell \pm \varkappa_{r} = \ell' \pm \varkappa_l$, the side peaks correspond to a change in photon momentum given by ${\delta \ell = \pm \Delta \ell}$ (recalling ${\Delta \ell = \varkappa_{r}-\varkappa_{l}}$). To understand the magnitude of the LO term compared to the NLO term, we can use the low energy expansion of the box diagram \cite{DeTollis:1964una,DeTollis:1965vna}. For the case of a plane wave background studied earlier, we found the LO terms were proportional to $\mathfrak{M}_{1212}-\mathfrak{M}_{1122} \sim s^2 - t^2 = 0$ and we expect the standing wave LO terms to scale similarly. For the kinematics of the side peaks scattered off the standing wave, one has $s = \pm 2 \ell \cdot \varkappa_{r}$ and $t = \mp 2 \ell \cdot \varkappa_{l}$, giving $s^2 - t^2 \sim \eta_{r}^{2}-\eta_{l}^2$. Analogously, the NLO term scales as $\sim s^3 - t^3 \sim \pm(\eta_{r}^{3}+\eta_{l}^{3})$. The NLO contribution thus remains dominant in the side peaks if 
\be
  \eta_{r}^{3}+\eta_{l}^{3} \gg \big|\eta_{r}^{2}-\eta_{l}^{2}\big| \; .
\ee

\section{Experimental Outlook}
One way to experimentally test the theory above is to pick a set-up of colliding beams in which the derivative term (NLO) is the main contribution to the experimental signature. For plane wave backgrounds, corrections to the NLO, \eqnref{eqn:CPflipProb1},  scale as $\chi^{2}$. To suppress these (and to avoid pair creation) one would require $\chi \ll 1$. Since measuring photon polarisation flip at LO is already an experimental challenge, the NLO signal should ideally not be much smaller than this. The ratio of the derivative term to the `standard' non-derivative HE term scales as $\eta$, and therefore it would lead to a significant NLO signal if $\eta \not \ll 1$. Since $\chi = \eta \xi$ in a plane wave, these two requirements then imply $\xi$ should not be too large. An example combination would be a high energy photon source with a weakly-focussed laser. To obtain some approximate numerical values, we model the envelope of the laser pulse as
\be
  \psi(\vphi) = \sin^{2}\left(\frac{\vphi}{2N}\right); \qquad 0 <
  \vphi<2\pi N,
\ee
and $\psi(\vphi)=0$ otherwise, such that $N$ measures the phase duration.  Some typical example parameters and corresponding polarisation flip probabilities, calculated from \eqref{eqn:CPflipProb1}, are given in \tabref{tab:expParams1}. Alternative scenarios at the electron-positron collider of the Future Circular Collider (FCC-ee) could include an inverse Compton scattering source of high energy photons with energies up to around $100\,\trm{GeV}$ \cite{Agapov:2928809}.

\begin{table}[h!!]
    \centering
    \begin{tabular}{ c | c | c | c | c }
    & $\eta$ & $\xi$ & $N$ & $\tsf{P}$\\
    \hline
     ELI-Beamlines \cite{Macleod:2023asi} & 0.01 & 20 & 40 & 
     $5.2\times 10^{-8}$\\
     LUXE \cite{Abramowicz:2021zja,LUXE:2023crk} & 0.1 & 5 & 16 & 
     $1.3\times 10^{-6}$\\
     FCC-ee \cite{Agapov:2928809} & 1 & 0.5 & 32 & 
     $2.6\times 10^{-6}$\\
\end{tabular}
\caption{Some example parameters and ensuing probabilities for the polarisation flip of an LP photon in a CP background (circular birefringence/optical rotation).}\label{tab:expParams1}
\end{table}

\section{Conclusion}
Vacuum \emph{circular} birefringence (or optical activity) is a prediction of QED that has been much less studied than its linear counterpart. It occurs whenever there is a non-zero probability for the linear polarisation of a probe photon to flip upon traversing a circularly polarised background. This endows the vacuum with a preferred `handedness' and thus turns it into a chiral `medium' akin to a solution of dextrose. 

Both variants of birefringence can be studied using a 2-by-2 S-matrix formalism with polarisation or helicity states which is basically equivalent to the Jones matrix approach of optics. The outcomes can be intuitively visualised on the Poincar\'e sphere (Figs. \ref{fig:PS1} and \ref{fig:PS2}). The flip amplitudes/S-matrix entries, though, must be calculated in QED or its low-energy effective approximation as has been done in this paper.

In this context it was already noted in \cite{King:2023eeo} that the vacuum circular birefringence signal cannot be described by the standard Heisenberg-Euler approach (the leading order in a derivative expansion). In this paper we have shown that the inclusion of derivative corrections reproduces the low-energy limit of the Volkov approach based on the dressed polarisation tensor. A third method of using the helicity amplitudes for photon-photon scattering in full QED leads to agreement with these other two methods.

Calculating the derivative corrections to the HE Lagrangian is a non-trivial task; even for the first correction, i.e.\ the next-to-leading-order Lagrangian, there are three different expressions stated in the literature. We outlined how Hilbert series can be used to identify the (minimal) number of inequivalent operators in the expansion and matched these operators with helicity amplitudes.

We also demonstrated how a derivative expansion can be performed to extend the locally constant field approximation (LCFA) for photon polarisation flipping. Although the LCFA usually only depends on the strong-field parameter, $\chi$, the extended LCFA with derivative expansion generated extra powers of $\eta$, thus doubling the dimensions of parameter space.  

A virtue of the effective field theory approach is its capability to tackle more realistic backgrounds with less symmetry than the super-integrable plane wave. We demonstrated this by calculating the polarisation flip of a linearly polarised photon scattering off a rotating standing wave background given by the superposition of \emph{two} circularly polarised plane waves.

The observation of vacuum circular birefringence or `vacuum chirality' would be a test of higher dimensional operators in the low-energy effective field theory for QED and experimentally yield the next-to-leading order effective couplings (collectively denoted $c_{4,2}$ in the Introduction. To realise this goal one has to achieve a sufficiently high centre-of-mass energy while maintaining a moderate strong-field parameter.

\begin{acknowledgements}
The authors acknowledge support from The Leverhulme Trust, Grant No. RPG-2024-142.
\end{acknowledgements}

\bibliographystyle{apsrev}
\bibliography{current}

\end{document}